# STRUCTURAL RELAXATION KINETICS FOR FIRST AND SECOND-ORDER PROCESSES: APPLICATION TO PURE AMORPHOUS SILICON

P. Roura* and J. Farjas

GRMT, Department of Physics, University of Girona, Campus Montilivi, E17071-Girona, Catalonia, Spain.

## Abstract

The structural relaxation of amorphous materials is described as arising from the superposition of elementary processes with varying activation energies. We show that it is possible to obtain the kinetic parameters of these processes from differential scanning calorimetry experiments. The transformation rate is predicted for the transient decay when an isotherm is reached and for the relaxation threshold detected in partially relaxed samples. Good agreement is obtained with experiment if the individual components transform through first-order kinetics, but inconsistencies arise for second-order components. Our analysis, that improves the classical treatment by Gibbs et al.[1], allows the activation energies and the pre-exponential rate constants to be extracted independently. When applied to a-Si, we conclude that the pre-exponential rate constant is far from constant. The kinetic parameters obtained from DSC are used to analyze the relaxation of a-Si in pulsed laser experiments and to discuss the relationship between structural relaxation and crystallization.

* Corresponding author, pere.roura@udg.es, Tel. 34972418383 fax. 34972418098

## Introduction

The structure of an amorphous material is metastable. During isothermal annealing, a moderate change of its properties is usually observed at short times (structural relaxation) followed by a very pronounced change at longer times (crystallization). This characteristic two-step evolution of amorphous materials can also be revealed by measuring the heat evolved when the material is heated at a constant rate [usually in a differential scanning calorimeter (DSC)],

$$dT/dt \equiv \beta. \qquad (1)$$

In the thermogram of Fig. 1 the power evolved, $\dot{Q}$, is plotted as a function of the temperature for hydrogenated amorphous silicon [2]. The weak unstructured signal detected at low temperatures is due to structural relaxation whereas the intense sharp peak corresponds to crystallization. Similar thermograms have been measured for other amorphous materials, such as ball milled silicon [3], metallic powders [4] and supercooled alloys [5, 6]. The characteristic unstructured signal of structural relaxation is also found in crystalline materials with a high concentration of non-equilibrium point or line defects: cold worked metals [7] and irradiated materials [8]. In these cases the low-temperature transformation is known as recovery.

In this paper, we will show that it is possible to obtain information about the kinetics of structural relaxation from thermograms similar to that of Fig. 1. The main difficulty lies in the fact that, in contrast to crystallization, structural relaxation does not occur around a peak temperature. When a structural transformation gives a well defined peak in the thermogram, its thermal activation is usually described by a reaction rate constant that follows an Arrhenius temperature dependence:

$$k_{Tp}(T) = \nu_{Tp} e^{-E_{Tp}/RT}, \qquad (2)$$

where $\nu_{Tp}$ is the pre-exponential constant, $E_{Tp}$ the activation energy and R the gas constant (the need for the subindex $T_p$ will become clear below). The usual way to determine $\nu_{Tp}$ and $E_{Tp}$ consists of measuring the peak temperature at different heating rates. The transformation rate tends to zero at low and at high temperatures because of the dependence of $k_{Tp}$ on T and because the transformation has reached completion, respectively. Consequently, when the kinetics is governed by a single activation energy, the reaction rate reaches a maximum value at an intermediate temperature, $T = T_p$, and gives a characteristic peak in the thermogram. The peak temperature increases with the heating rate and, for most kinetic models, the Kissinger equation [9]:

$$\frac{E_{Tp}}{RT_p^{\ 2}} = \frac{\nu_{Tp} e^{-E_{Tp}/RT_p}}{\beta} \equiv \frac{k_p}{\beta} \qquad (3)$$

is accurate enough in a large range of conditions and is used to extract $\nu_{Tp}$ and $E_{Tp}$ from the experiment. In Eq. (3) $k_p$ is the value of $k_{Tp}(T)$ at the maximum of the transformation peak.

Owing to its structural disorder, a distribution of atomic configurations coexists in any amorphous material. Consequently, we will consider that a superposition of independent microscopic processes with different activation energies contributes to the structural relaxation [1, 10]. In the inset of Fig. 2 this assumption is applied to explain the calorimetric signal. Every microscopic component contributes with a peak centered approximately at a value of $T = T_p$ determined by $\nu_{Tp}$ and $E_{Tp}$ through the Kissinger Eq. (3). Therefore, the thermogram can be calculated as the convolution of the peak temperature distribution, $n(T_p)$, by the "peak shape", $R_{Tp}$ (T):

$$\dot{Q}(T) = V\int_0^{\infty} h(T_p) n(T_p) R_{Tp}(T) dT_p \,, \qquad (4)$$

where V is the material volume; $n(T_p)dT_p$ is the number of states (or defects) per unit volume with peak temperatures between $T_p$ and $T_p+dT_p$ and $h(T_p)$ the heat evolved when one of these states is transformed (specific enthalpy). The "peak shape" is the probability per unit time that a given microscopic state is transformed when the material is heated at constant rate. The condition that any state will be transformed at the end of the heating ramp imposes the normalization condition:

$$\int_0^{\infty} R_{Tp}(T) dT = \beta \,. \qquad (5)$$

The integration of Eq. (4) is straightforward if $h(T_p)n(T_p)$ is a slowly varying function and we obtain:

$$\dot{Q}(T)/\beta \approx V\, h(T)\, n(T) \,, \qquad (6)$$

i.e., the calorimetric signal is a measure of the peak temperature distribution (in fact, the activation energy distribution) modulated by the specific enthalpy.

In Section II.1 we will show that it is possible to determine the $\nu_{Tp}$ and $E_{Tp}$ values of a particular individual component by analyzing the decay of the calorimetric signal when an isotherm at $T = T_p$ is reached after a heating ramp. This means that, in contrast with many analyses [11, 12] where $\nu_{Tp}$ is taken as a constant for any component and $E_{Tp}$ is obtained from the relaxation rate at a given temperature through Eq. (2), we will obtain both $\nu_{Tp}$ and $E_{Tp}$ from the experiment. The reliability of our results will be tested in Section II.2 with the analysis of a complementary experiment consisting of a heating ramp after an isothermal annealing of the material. In this case, the shift of the signal threshold with the annealing time will deliver $\nu_{Tp}$ and $E_{Tp}$.

The degree of relaxation can be quantified through the total heat evolved,

$$\Delta P = \int_0^{t_f} \dot{Q}\, dt\,, \tag{7}$$

where $t_f$ is the duration of the anneal. Gibbs et al. [1] demonstrated that, during isothermal annealing, $\Delta P$ is proportional to RT·Ln(t). This prediction has been experimentally observed in a variety of materials for which a given property follows the Ln(t) dependence [1, 11, 13, 14, 15, 16]. However, the proportionality with RT is not obeyed; on the contrary, the slope of the curve "Property vs Ln(t)" usually varies slowly [11] or is independent [14] of T. As we will point out in Section II.1, during an isotherm the relaxation rate, $\dot{Q}$, depends on the heating rate, β, of the previous heating ramp [17]. Our approach allows us to show clearly under which conditions the proportionality with RT holds. In addition, by including the heating ramp, the divergence of the relaxation degree at t = 0 inherent in the logarithmic dependence is solved (Section II.3).

The accurate predictions developed in this paper have been possible because simple analytical functions describing the peak shape [$R_{Tp}(T)$ in Eq. (4)] have recently been published [18]. We have carried out all the analyses for first and second-order elementary processes. For a first-order process, the transformation rate is described by the equation:

$$\frac{d\alpha_{Tp}}{dt} = k_{Tp}(T)(1-\alpha_{Tp})\,, \tag{8}$$

where $\alpha_{Tp}$ is the transformed fraction ($0 < \alpha_{Tp} < 1$). For a constant heating rate, the exact solution of Eq. (8) can be approximated with good accuracy by [18]:

$$\alpha_{Tp}(T) = 1 - \exp(-e^{u_p})\,, \tag{9}$$

where $T_p$ is the temperature where $d\alpha_{Tp}/dt$ is at its maximum and

$$u_p \equiv \frac{E_{Tp}}{RT_p}\frac{T-T_p}{T_p} = \frac{k_p}{\beta}(T-T_p)\,, \tag{10}$$

where the last term is obtained by applying the Kissinger Eq. (3). The time derivative of Eq. (9) leads us to the peak shape for a first-order reaction:

$$\frac{d\alpha_{Tp}}{dt} = k_p e^{u_p}\exp(-e^{u_p}) \equiv R_{Tp}(T)\,. \tag{11}$$

It is necessary to mention here that this analytical solution has been obtained by substituting $k_{Tp}(T)$ in Eq. (8) by its approximate value:

$$k_{Tp}(T) \approx k_p e^{u_p} , \quad (12)$$

which is obtained by linearizing the exponent around the peak temperature $T_p$. This approximation will be used to derive most of the relationships in this paper.

For a second-order process,

$$\frac{d\alpha_{Tp}}{dt} = k_{Tp}(T)(1-\alpha_{Tp})^2 , \quad (13)$$

the corresponding approximate solution is [18]:

$$\alpha_{Tp}(T) = \frac{e^{u_p}}{e^{u_p}+1} , \quad (14)$$

and the peak shape

$$\frac{d\alpha_{Tp}}{dt} = \frac{k_p e^{u_p}}{(e^{u_p}+1)^2} \equiv R_{Tp}(T) . \quad (15)$$

In Fig. 2 the peaks of first- and second-order processes are compared. Despite the pronounced differences between them, it will be shown in Section II.1 that the signal decay during an isotherm is the same for both cases. In addition, in Section II.2 we will show that the shape of the relaxation threshold after annealing is also similar. Although these results imply that it is difficult to determine the reaction order from these kinds of experiments, in Section II.4 it will be argued that, from a formal point of view, a superposition of independent second-order components is inconsistent. This conclusion contrasts with a number of authors' claim that structural relaxation in amorphous silicon is due to second-order recombination processes [19].

The main results obtained in Section II will be applied in Section III to discuss several aspects of structural relaxation in amorphous silicon.

## II. Theory

### II.1 Isothermal relaxation rate after heating at a constant rate

During a heating ramp, the convolution integral of Eq. (4) holds. This means that a given component will relax in a temperature interval around its peak temperature $T_p$. If the temperature rise does not stop, its transformation rate $\dot{\alpha}_{Tp}$ will tend to zero once $T_p$ has been surpassed, according to its characteristic peak shape. However, this evolution will change drastically when an isothermal period at $T = T_a$ is reached (subindex "a" stands for "annealing"). The evolution for first-order kinetics is drawn in Fig. 3. At the beginning of the isotherm ($t = 0$) the transformed fraction $\alpha_{Tp}(T_a)$ is given by Eq. (9) with $T = T_a$. This value is required to solve the transformation rate Eq. (8) which delivers the time dependence of $\alpha_{Tp}$ during the isotherm:

$$\left[1-\alpha_{Tp}(t)\right]=\left[1-\alpha_{Tp}(T_a)\right]e^{-k_{Tp}(T_a)t} \tag{16}$$

and its transformation rate:

$$\dot{\alpha}_{Tp}(t)=k_{Tp}(T_a)\left[1-\alpha_{Tp}(t)\right]. \tag{17}$$

The relaxation rate can now be calculated by integrating all the components ($0 < T_p < +\infty$):

$$\dot{Q}(t)=V\int_{T_p=0}^{+\infty}k_{Tp}(T_a)\left[1-\alpha_{Tp}(t)\right]h(T_p)n(T_p)\,dT_p\approx V\,h(T_a)n(T_a)\int_{T_p=0}^{+\infty}k_{Tp}(T_a)\left[1-\alpha_{Tp}(t)\right]dT_p \tag{18}$$

where the approximate expression is correct if $n(T_p)h(T_p)$ is a slowly varying function. The integral does not have an exact analytical solution unless we use the approximate value for $k_{Tp}(T)$ of Eq. (12) and $k_p$ is substituted by an average value, $\overline{k_p}$. Changing the integration variable to

$$u\equiv\frac{k_p}{\beta}(T_a-T_p) \tag{19}$$

leads to:

$$\dot{Q}(t)=Vh(T_a)n(T_a)\beta\int_{-\infty}^{k_pT_a/\beta}e^u\exp\left[-e^u\right]\exp\left[-\overline{k_p}te^u\right]du, \tag{20}$$

where, according to the Kissinger equation, the upper integration limit is equal to $E_{Ta}/RT_a$ and, thanks to the sharp dependence of the integrand on u, it can be substituted by $+\infty$, because for most solid state transformations $60 > E_{Ta}/RT_a > 10$ [20] (even for $E_{Ta}/RT_a$ as low as 5, the relative error is below $10^{-60}$). We finally obtain the relaxation rate:

$$\dot{Q}(T_a,t)=\frac{V\beta h(T_a)n(T_a)}{\overline{k_p}t+1}. \tag{21}$$

Under the same approximations used so far, an identical result is obtained when the calculation is carried out for a superposition of second-order processes.

Note that the value at $t = 0$ is just the DSC signal during the heating ramp [Eq. (6)] as it should be. The time dependence of the transient is governed by a single parameter $\overline{k_p}$ whose value is approximately [Eq. (3)]:

$$\overline{k_p}\approx\frac{\beta E_{Ta}}{RT_a^2}, \tag{22}$$

where $E_{Ta}$ means the activation energy of the component with peak temperature $T_p = T_a$. Since, for a given relaxation process, the range of activation energies is restricted within a maximum variation of a factor of 5 (say from hundreds of meV to several eV) and $E_{Ta}$ increases with $T_a$, Eq. (22) tells us that the decay rate of the relaxation signal will have a smooth dependence on the annealing temperature.

The decay of the DSC signal predicted by Eq. (21) has been compared with the experiment in Fig. 4. The measurements were done on c-Si amorphized by low-temperature ion implantation by Roorda et al. [19]. Three identical samples were heated at 40 K/min from room temperature until the annealing temperature was reached ($T_a$ = 200, 350 and 500ºC). Good fits for the three transients have been reached. The slight discrepancies observed for long times, where the signal becomes small, could be related to slight apparatus instabilities or they could reflect a departure from our simple model. A slow monotonous diminution of $\overline{k_p}$ is obtained as $T_a$ is increased: 0.055 and 0.050 $s^{-1}$ for $T_a$ = 200 and 500ºC, respectively. Applying the approximate Kissinger Eq. (22) allows us to obtain the (average) activation energy of the components relaxing around $T_a$ and, finally, substituting T for $T_a$ and $k_{Tp}(T)$ for $\overline{k_p}$ in Eq. (2) delivers the pre-exponential factor, $\nu_{Ta}$. These values are detailed in Fig. 4. Note that $\nu_{Ta}$ is far from being constant.

In the solution given by Eq. (21), $\overline{k_p}$ is kept constant independent of t. However, the experimental results of Fig. 4 indicate that $\overline{k_p}$ depends on the annealing temperature. In fact, the relaxation rate at very long annealing times is due to the transformation of the high temperature components which transform with a lower value of $\overline{k_p}$. Here we will show that, for the transients of Fig. 4, $\overline{k_p}$ is almost constant, time independent. For this, it is necessary to determine the relative contribution of each component to $\dot{Q}$ at various elapsed times. This contribution is simply the value of the integrand in Eq. (18). For the particular transient at 350ºC and for first-order processes, the result is the bell-shaped curves shown in Fig. 5. Note that, at the beginning of the transient (t = 0), the maximum contribution is that of the component with $T_p = T_a$ whereas the relative contribution of the high temperature components increases with time. We see that, at the end of the transient (at 300 s), the maximum has shifted around 30ºC above $T_a$. For the particular case of a-Si analyzed here, this shift would represent a negligible variation of $\overline{k_p}$ lower than 1 %. This result is quite unexpected because, despite of the contribution of a broad distribution of energy barriers (approximately from 2.8 to 3.0 eV for the transient at 350ºC), a simple kinetic constant ($\overline{k_p}$) is enough to describe the decay of the relaxation rate.

## II.2 Threshold of the relaxation rate after an isothermal annealing

An experiment that can deliver the same information as the previous experiment [i.e. $\overline{k_p}$ and $n(T_p)h(T_p)$] consists of measuring the heat evolved at a constant heating rate from a sample that has been partially relaxed by heating it at the same rate and held at the same temperature $T_a$ for $t = t_a$. If the second ramp begins at a low enough temperature then a threshold of the DSC signal will be measured at a temperature higher than $T_a$. This behavior can be easily understood with the help of Fig. 5 where the sigmoidal curves are the untransformed fractions [$1-\alpha_{Tp}(T_a,t_a)$]. For longer annealing times, the components that remain untransformed have higher peak temperatures. Consequently, the relaxation signal measured in a subsequent heating ramp will have a threshold at a temperature higher than $T_a$.

Since the $\dot{Q}(T)$ signal during a heating ramp is the convolution of the [$1-\alpha_{Tp}(T_a,t_a)$] curves by the peak-shape, according to Eq. (4), calculating the $\dot{Q}(T)$ threshold is straightforward for first-order kinetics:

$$\dot{Q}(T) = V\, h(T)n(T)\int_0^{\infty}\left[1-\alpha_{Tp}(T_a,t_a)\right]R_{Tp}(T)dT_p\,, \qquad (23)$$

where $\left[1-\alpha_{Tp}(T_a,t_a)\right]$ is given in Eq. (16) with $t = t_a$, i.e.:

$$\left[1-\alpha_{Tp}(T_a,t_a)\right] = \exp\left[-e^{\frac{k_p}{\beta}(T_a-T_p)}\right]\exp\left[-k_p t_a e^{\frac{k_p}{\beta}(T_a-T_p)}\right] \qquad (24)$$

and $R_{Tp}(T)$ is given in Eq. (11). No approximations are needed to carry out the integration with the following lengthy result:

$$\dot{Q}(T) = \frac{\beta h(T)n(T)}{(1+\overline{k_p}t_a)e^{-u_a}+1}\left[1-\exp\left(-e^{\frac{\overline{k_p}}{\beta}T}(\overline{k_p}t_a e^{-u_a}-e^{-u_a}+1)\right)\right] \approx \frac{\beta h(T)n(T)}{(1+\overline{k_p}t_a)e^{-u_a}+1}\,, \qquad (25)$$

where the factor inside the brackets has been neglected because $\exp(\overline{k_p}\,T/\beta) = \exp(E_{Tp}T/RT_p^2) >> 1$ and

$$u_a \equiv \frac{\overline{k_p}}{\beta}(T-T_a)\,. \qquad (26)$$

Finally, the shift of the relaxation threshold can be easily quantified from Eq. (25) by calculating the temperature at which $\dot{Q}$ reaches half its maximum value:

$$T_{1/2} = T_a + \frac{\beta}{\overline{k_p}}Ln(1+\overline{k_p}t_a)\,. \qquad (27)$$

For a given value of $T_a$ and $t_a$, the shape of the threshold can be easily calculated by introducing the value of the only free parameter, $\overline{k_p}$, that is, the average of $k_p$ over the components that contribute to the threshold. We have fitted the corresponding experiments of Roorda et al. [19] who annealed the a-Si samples during 45 min at 150, 230 and 300ºC after heating them at 40 K/min. The calculated curves fit the experimental points quite well (Fig. 6) and the values of $\overline{k_p}$ thus obtained are close to those obtained in the previous section (inset of Fig. 6). We do not know the reason for the observed small discrepancies.

Again, the calculations have been carried out for second-order kinetics (see Appendix A). Although the fit to the experimental points are reasonably good, the values of $\overline{k_p}$ thus obtained are much higher than those extracted from the isothermal decay (inset of Fig. 6). This discrepancy indicates that, probably, the relaxation signal of a-Si is not the superposition of second-order components.

**II.3 The degree of relaxation after isothermal annealing**

To interpret the evolution due to structural relaxation of certain material properties during isothermal annealing, the classical model by Gibbs et al. [1] assumes that 1) the transformation of any microscopic component makes an equal contribution to the property change, ΔP, independently of its activation energy. In addition, 2) this model simplifies the integration of the components by substituting the smooth evolution of $\alpha_{Tp}(T)$ (Fig. 2) by a step function and 3) considers that the preexponential constant is the same for any component. This model leads to a logarithmic evolution of ΔP:

$$\Delta P \propto A + RT_a Ln(t) . \qquad (28)$$

According to this solution, the slope of ΔP versus Ln(t) should be proportional to $T_a$.

With our formalism, we can also predict ΔP. The first assumption of Gibbs et al. is equivalent to considering that $Vh(T_p)n(T_p)$ is constant during the isotherm. Consequently, with this assumption, ΔP is formally equivalent to the "degree of relaxation" defined in Eq. (7). We can calculate ΔP through the integration of $\dot{Q}$ [Eq. (21)]:

$$\Delta P(t,\beta) = \Delta P(0,\beta) + C\frac{\beta}{\overline{k_p}} Ln(\overline{k_p}t+1) = \Delta P(0,\beta) + C\frac{RT_a^2}{E_{Ta}} Ln(\overline{k_p}t+1) , \qquad (29)$$

where C is constant, ΔP(0,β) is the relaxation degree just after the heating ramp and the last expression is obtained by using Eq. (3). For the particular case where $\nu_{Tp}$ is independent of $T_p$, it can be demonstrated that $T_p$ is proportional to $E_{Tp}$ [20] and the temperature dependence of Gibbs [Eq.(28)] is obtained. However, in general $\nu_{Tp}$ will depend on the particular component and, consequently, different temperature dependencies are expected. For the particular case of a-Si, the experiments analyzed in Sections II.1 and II.2 indicate that $\overline{k_p}$ is almost constant, independent of $T_a$ (inset of Fig. 6). Consequently, our Eq. (29) predicts that, for $\overline{k_p}$ t>>1, the slope of ΔP versus Ln($t_a$) (i.e. $\beta / \overline{k_p}$) should be almost independent of the annealing temperature. Measurements of the electrical conductivity evolution at different temperatures made by Shin et al. [11] confirm our prediction (the slope only doubles its value from 77 to 573 K).

Concerning the argument of the Ln function, our solution tends to that of Gibbs when $\overline{k_p}$ t>>1. In fact, Gibbs calculated the relaxation of a material that at t = 0 is already at the annealing temperature with all its components still untransformed. In our formalism, this corresponds to taking [1-$\alpha_{Tp}(T_a)$] = 1 in Eq. (16). With this change, the calculated DSC signal is proportional to 1/t and $\Delta P \propto Ln(t)$. Consequently, the Ln($t_a$) dependence of the Gibbs solution can be understood as a) the asymptotic behavior of our more general solution for long annealing times (i.e., for a time that is long enough for the initial conditions or the particular heating rate to be irrelevant) or, alternatively, as b) the solution for an instantaneous heating ramp. Since most isothermal annealing experiments are carried out after heating at a finite rate (like those in Section II.1) our solution represents a clear improvement with respect to that of Gibbs. In particular, it allows the kinetic parameter $\overline{k_p}$ to be obtained from the evolution of an appropriate property during the annealing. This important information is completely lost for the conditions in which the Gibbs solution is valid.

**II.4 Formal argument against a distribution of independent second-order processes**

We have seen in Section II.2 that the fits of the experimental threshold to a superposition of second-order components deliver unreliable values of $k_p$. However, since the work by Roorda et al. [19], many other authors [11, 12, 21] have assumed that the structural relaxation of a-Si is due to second-order components. The underlying microscopic mechanism is that structural relaxation was driven by the recombination of structural defects. Here we will show that, from a formal point of view, a superposition of second-order processes is not possible.

Consider a crystalline material with N identical defects per unit volume. Their recombination will be described through second-order kinetics, i.e.:

$$\frac{dN}{dt} = -k_N N^2 , \qquad (30)$$

where $k_N$ is the recombination probability and is given in $s^{-1}m^3$. This probability is independent of the initial defect concentration $N_0$ and, for a crystalline material, has a well defined activation energy value.

For an amorphous material, a continuum of activation energies arises and the state of the material will be described by the distribution of defects along their activation energy, n(E). If the recombination were restricted to defects with identical activation energy, the recombination rate would be negligibly small. This is so, because N in Eq. (30) should be replaced by n(E)dE (i.e. the concentration of defects around E) and, consequently, $dn(E)/dt \propto dE$. In other words, defects with different activation energies ($dE \neq 0$) should mutually recombine to obtain a finite recombination rate. However, this condition contradicts one of the basic assumptions of our description of the relaxation process. Namely, the components with different activation energies must relax independently. We conclude that, although recombination of defects does occur in amorphous materials, it cannot be described by a superposition of second-order processes.

**II.5 Kinetics of the mutual recombination of structural defects**

When defects with different activation energies mutually recombine, their recombination kinetics is far from being simple (see, for instance, the attempt developed in ref. [22]). Here we will show that, in some instances, defect recombination can be described by a superposition of first-order processes. This could be the case of dangling-bonds (DB) in a-Si. Consider a pair of DB with very different activation energies for diffusion. At any temperature, the DB with a lower activation energy will follow a random path while the other DB will remain essentially at a fixed site. This implies that their mutual recombination will be governed by the lowest activation energy. Now, take N dangling bonds distributed along a continuum of activation energies. Owing to their higher mobility, the DBs with the lowest activation energies will recombine first, according to a first order process, i.e.:

$$\frac{dn(E)}{dt} = -k_N N \cdot n(E) \quad . \qquad (31)$$

An interesting feature of this model is that the distribution of defects along their activation energy [n(E) in Eq. (31)] is an ‘apparent’ distribution related in a simple way with the real distribution just before relaxation begins (see Appendix B). The first-order equation (31) is obtained under the simplification that the only mobile defects are those with the minimum activation energy. However, there exists a finite range of energies above this minimum value with defects mobile enough to contribute appreciably to recombination. This simplification could explain the observed discrepancies between experiment and theory (Figs. 4 and inset of Fig. 6).

## III. Discussion: structural relaxation of a-Si

### III.1 Defect recombination and bond-angle strain

Since calorimetry was first used to detect structural relaxation in pure a-Si [23], many papers have been published to elucidate the main microscopic changes and mechanisms involved. In particular, the role of defect annihilation has been stressed [11, 12, 19, 21, 24] and most authors have assumed a superposition of bimolecular second-order components. However, it is well known that structural relaxation of a-Si also involves reducing bond-angle strain, as revealed by Raman spectroscopy [25, 19] and microscopic models of a-Si [26, 27]. Thus, the question arises concerning the relationship between defect recombination and bond-angle strain reduction and the contribution of both processes to the heat evolved during relaxation.

A critical review of the literature [28] has led us to conclude that the heat of relaxation is mainly due to the reduction of bond-angle strain and that defect annihilation only makes a minor contribution. This conclusion is further reinforced by the results of several authors summarized in Fig. 7. Coffa et al. [12] and Shin et al. [11] used electrical conductivity measurements to deduce the number of structural defects that recombined at a given temperature (in Fig. 7, $E_F$ means that we refer to the defects located near the Fermi level). In particular, Coffa et al. [12] predicted high heat release during relaxation from 100 K to room temperature due to defect recombination. The DSC experiments by Mercure et al. [21] gave an unexpectedly flat signal in this temperature range. In addition, although both the DSC signal and defect recombination diminish above room temperature, the decrease is much lower for the calorimetric signal (Fig. 7). According to Eq. (6), this means that the heat released per defect [h(T)] increases with the temperature. In this section we will solve the apparent contradiction between the calorimetric and electrical results.

We assume that the DSC signal is mainly due to the reduction of bond-angle strain [28]. Without the contribution of structural defects, bond rearrangement by thermal annealing leading to lower bond-angle strain is very difficult because it would entail the simultaneous bond-breaking and rearrangement of a large number of bonds. However, the process is much easier when the bonds break and rearrange due to the random path of the structural defects. Therefore, the kinetics of structural relaxation is governed by defect recombination. According to Sections II.4 and II.5, structural defects probably recombine following first-order kinetics. However, the important fact is that, as the temperature increases, the remaining density of defects [N in Eq.(31)] diminishes and, consequently, the path followed by any defect until it recombines increases. This provides more opportunities for the Si-Si network to rearrange. Therefore, in agreement with the experiment (Fig. 7), the heat released per defect will increase with temperature. On the other hand, since the power released by bond rearrangement is proportional to the number of defects that change their position per unit time, which is in turn proportional to the number of mobile defects [n(E) in Eq. (31)], then the calorimetric signal will be the superposition of first-order processes [Eq. (31)].

This model agrees with the correlation found by Stolk et al. between the bond-angle dispersion and the concentration of defects after isothermal annealing [24]. These authors implicitly assumed that recombination of defects follows first-order kinetics.

**III.2 The kinetic parameters of a-Si relaxation**

In Fig. 8 we summarize the activation energies, $E_{Tp}$, and pre-exponential constant rates, $\nu_{Tp}$, obtained by fitting the isothermal DSC transients (Fig. 4) and the relaxation threshold measured at 40 K/min by conventional DSC [19] and at $\approx 4\ 10^4$ K/s by nanocalorimetry [21]. Despite the error bars and the dispersion of points, it is clear that $\nu_{Tp}$ is not independent of $E_{Tp}$. This is in contrast with all the analyses concerning a-Si published so far, where a constant value of $10^{13}$ $s^{-1}$ was assumed. The exponential increase of the preexponential factor with the activation energy is usually known as the Meyer-Neldel or compensation effect. Although its microscopic origin is under debate [29], it is formally interpreted as arising from an entropic barrier, $\Delta S$:

$$\nu_{Tp} = \nu_0 e^{\Delta S / R} \tag{32}$$

where $\nu_0$ would be the atomic vibration frequency ($\approx 10^{13} s^{-1}$ for Si). The values of $\Delta S$ thus deduced are detailed in the right hand axis of Fig. 8.

The values of Fig. 8 allow the distribution of activation energies, $G(E_{Tp})$, to be calculated from the defects that recombine at a given temperature, $G(T_p)$:

$$G(E_{Tp}) = G(T_p) \frac{dT_p}{dE_{Tp}} . \tag{33}$$

Although, as pointed out by several authors [11, 12], $dT_p/dE_{Tp}$ would only change slightly if $\nu_{Tp}$ were increased by several orders of magnitude, this is not the case for the range of activation energies. Our results indicate that at ≈500ºC, relaxation is governed by components with activation energies as high as 3.5 eV, whereas an upper limit of 2.5-2.8 eV was obtained with $\nu_{Tp} = 10^{13}$ $s^{-1}$ [11, 12].

In addition, if the defect annihilation mechanism proposed in Section II.5 is valid, the measured $G(E_{Tp})$ distribution has to be interpreted as the "apparent" distribution because the $E_{Tp}$ component entails the mutual recombination of the defects with this energy and defects with higher activation energies (see Appendix B and its Fig. 10 for details).

### III.3 Structural relaxation under pulsed laser annealing

We will show here that the kinetic parameters of Fig. 8 are still valid when relaxation is induced at the very high heating rates achieved with pulsed lasers. Grimaldi et al. [30] measured the dependence of the melting temperature of a-Si on the relaxation degree. Before laser irradiation, one sample was relaxed by thermal annealing at 450ºC during 60 min and another sample was left unrelaxed. When these samples were irradiated with pulses of 20 ps, the unrelaxed sample melted at a lower pulse energy, which indicates that its melting temperature is around 160 K lower than that of the relaxed sample. This result confirmed the theoretical predictions [31] and indicates that structural relaxation did not occur during laser heating. In contrast, nearly the same melting temperature was deduced for both samples with 30 ns laser pulses, which indicates that, under these conditions, relaxation occurred during laser heating.

In Fig. 9 we sketch our predictions concerning these experiments. We have plotted the peak temperature of the components that, at 40 K/min, relax at 100, 300 and 500ºC that (according to Fig. 8 and the Kissinger equation) correspond to the following pairs of ($\nu_{Tp}$, $E_{Tp}$) values: (5 $10^{16}$ $s^{-1}$, 1.3 eV), (2 $10^{18}$, 2.2) and (5 $10^{22}$, 3.7). These components are referred as the "initial", "intermediate" and "final" stages of relaxation. According to the Kissinger Eq. (3), $T_p$ increases with the heating rate. This evolution of the three components is compared in Fig. 9 with the melting temperature of unrelaxed a-Si, $T_M$. Since $T_M$ is around 1300 K, the heating rate achieved with laser pulses will be around $100/\Delta t < \beta < 1000/\Delta t$ (K/s), where $\Delta t$ is the pulse duration. When the energy per pulse is increased, the maximum temperature achieved is higher; however, the heating rate remains essentially the same. From Fig. 9 it is clear that, for picosecond laser pulses, $T_M$ is reached before the material has relaxed appreciably. However, when $T_M$ is reached with nanosecond laser pulses, relaxation is almost complete. This prediction agrees with the results by Grimaldi et al. [30] and also with those of Stolk et al. [24] who observed, for nanosecond pulses, a progressive relaxation as the laser energy increased.

In contrast with the conclusions drawn from the early relaxation experiments with lasers [32], our analysis clearly shows that the relaxation kinetics is essentially the same for heating rates spanning many orders of magnitude, $1 \approx \beta < 10^{14}$ K/s.

### III.4 Structural relaxation and crystallization

In contrast with structural relaxation, crystallization is discontinuous: there is a large finite energy difference between the initial (a-Si) and final state (c-Si) and it occurs heterogeneously with an abrupt interface between them. Its kinetics is governed by the nucleation and growth rate constants, with well-defined activation energies. In addition, an incubation time for nucleation is usually observed [33, 34]. Since the initial state for crystallization is a particular configuration of a-Si reached by structural relaxation, it seems natural to question the influence of structural relaxation on crystallization.

Very accurate measurements of the growth rate, $v_g$, have been made by monitoring the epitaxial crystallization of amorphous layers obtained in c-Si [35, 36]. Lu et al. [36] explicitly addressed the question of possible changes in $v_g$ in relation to different relaxation degrees. One relaxed sample and one unrelaxed sample were heated in about 1 s to the crystallization temperature of 630ºC. Since the same value of $v_g$ was measured in both samples, the authors concluded that $v_g$ was independent of the relaxation degree. However, an analysis similar to that given in Section III.2 reveals that both samples reached the crystallization temperature in an almost fully relaxed state. The same conclusion in reached when analyzing the impressive experiments by Olson et al. [35], covering a variation of $v_g$ of 10 orders of magnitude. In these experiments, the material is fully relaxed after less than 1/1000 of the isotherm duration. Thus, we conclude that, as far as we know, all values of $v_g$ reported so far are related to the crystallization of fully relaxed a-Si. In fact, since $v_g$ is very slow compared to the relaxation rate, we think it is very difficult to observe the epitaxial crystallization of partially-relaxed a-Si.

Concerning the nucleation rate, we do not know of any experiment devoted to analyzing the dependence of the nucleation rate on relaxation. Here we will show that the material is fully relaxed after the incubation time, $t_{inc}$. $t_{inc}$ was measured in the 580-780ºC range [33, 36]. At 580ºC, $t_{inc} \approx 10^5$ s, whereas the time needed to relax the component with a peak temperature of 500ºC (at 40K/min) was around 0.1 s. At 780ºC, $t_{relax}$ ($\approx 10^{-5}$ s) is still much shorter than the incubation time ($\approx$ 100 s). These calculations show that the material has "plenty of time" to reach a highly relaxed state before crystallization (nucleation) begins.

This last conclusion raises a very interesting question. Microscopic models of a-Si [26] cannot be built with bond-angle dispersion values below $\Delta\theta \approx 7^{\circ}$. Below this value, the models are intrinsically unstable and transform quickly into c-Si. It therefore seems reasonable to think that the relaxed states of a-Si reached experimentally will approach this value. Raman spectroscopy confirms that, at the onset of crystallization, the bond-angle dispersion falls around the theoretical value [19, 24, 28]. What is surprising is the fact that relaxation of the bond-angle strain stops at this particular value. Apparently, nothing occurs during the time elapsed from the end of the observable effects of relaxation until crystallization begins. It seems that the configuration with $\Delta\theta \approx 7^o$ is highly stable against relaxation. Since relaxation of $\Delta\theta$ depends on the random path of structural defects, one possible explanation would imply that the remaining defects detected after the relaxation experiments [24] are not mobile. Another possible explanation would be more intrinsic to the structure of a-Si and would imply, as microscopic modeling suggests, that no configuration of a-Si exists below $\Delta\theta \approx 7^o$, which makes further structural relaxation impossible.

**Conclusions**

We have shown that it is possible to extract the kinetic parameters of structural relaxation from two simple DSC experiments. Although emphasis has been put in DSC, our results can be also applied to any technique able to determine the reaction rate from the evolution of any material property that evolves with structural relaxation.

**Acknowledgments**

This work was partially supported by the Spanish Programa Nacional de Materiales under contract number MAT2006-11144 and by the Generalitat de Catalunya under contract number 2005SGR-00666.

**Appendix A: Relaxation threshold for second-order components**

Calculating the relaxation threshold for second-order components is more cumbersome than the isothermal decay because, unlike first-order processes, the temperature peak of a particular component that has been partially transformed during annealing at $T_a$ is shifted to a higher temperature. The new peak temperature $T_p$' is related, with good accuracy, to the initial peak temperature $T_p$ by the equation [20]:

$$\frac{1}{T_p'}-\frac{1}{T_p}\approx\frac{1}{T_p}\frac{Ln(1-\alpha_{Tp}(t_a))}{2+(E_{Tp}/RT_p)}\ . \qquad \text{(A4)}$$

where $[1-\alpha_{Tp}(t_a)]$ is the untransformed fraction after isothermal annealing and is given by Eq. (A1). The peak shape of this component during the second heating ramp $R_{Tp}'(T)$ is the second-order peak shape $R_{Tp}(T)$ of Eq. (15) with $T_p$ replaced by $T_p$':

$$R_{Tp}'(T)=\frac{k'_p\, e^{u'_p}}{(e^{u'_p}+1)^2}\,, \qquad \text{(A5)}$$

where $k'_p = k_{Tp}(T=T'_p)$.

Now, the shape of the relaxation threshold can be calculated by integrating the contribution of all the $T_p$ components:

$$\dot{Q}(T) = V\,h(T)n(T)\int_0^\infty \left[1-\alpha_{Tp}(T_a,t_a)\right]R_{Tp}'(T)dT_p \,. \qquad \text{(A6)}$$

Unlike Eq. (21) we have not been able to solve it analytically and, consequently, we have solved it numerically for several values of T. The result is plotted in Fig. 6.

**Appendix B.- Evolution of the distribution of structural defects during their mutual recombination**

Consider that n(E) is the initial distribution of defects. If the recombination probability is independent of the particular activation energies of the defects involved, then at $T = T_p$ the distribution will be, approximately (Fig. 10):

$$\begin{aligned} n_{ETp}(E) &= 0 & E < E_{Tp} \\ n_{ETp}(E) &= \frac{f(E_{Tp})}{f(E_{\min})}n(E) & E \geq E_{Tp} \end{aligned} \,. \qquad \text{(B1)}$$

When all defects with energy between $E_{Tp}$ and $E_{Tp}+\delta E_{Tp}$ disappear, an equal number of defects with $E_{Tp} > E$ will recombine with them (see Fig. 10):

$$n_{ETp}(E_{Tp})\cdot\delta E_{Tp} = -\int_{E_{Tp}}^{E_{\max}} \delta n_{ETp}(E)dE \,. \qquad \text{(B2)}$$

Since, like $n_{ETp}(E)$, $\delta n_{ETp}(E)$ is proportional to n(E) [Eq. (B1)], Eqs. (B1) and (B2) lead to:

$$f(E_{Tp})n(E_{Tp})\delta E_{Tp} = -\delta f(E_{Tp})\int_{E_{Tp}}^{E_{\max}} n(E)dE \equiv -\delta f(E_{Tp})I(E_{Tp}) \,, \qquad \text{(B3)}$$

Integration of Eq. (B3) with the initial condition $f(E_{Tp} = E_{min}) = f(E_{min})$, and use of Eq. (B1) leads to the desired result:

$$n_{ETp}(E_{Tp}) = n(E_{Tp})I(E_{Tp})/N_0 \,. \qquad \text{(B4)}$$

where $N_0$ is the initial defect density, $N_0 = I(E_{Tp} = E_{min})$. I.e. the density of defects that recombine following a process governed by $E_{Tp}$ is $2n_{ETp}(E_{Tp})$ and not $n(E_{Tp})$. We can thus define an apparent distribution as:

$$n_{ap}(E) \equiv 2n_E(E) = 2n(E)\int_E^{E_{\max}} n(s)ds/N_0 \,, \qquad \text{(B6)}$$

plotted in Fig. 10 as a dashed curve. $n_{ap}(E)$ is measured by experiments and leads to a relaxation process formally described by a superposition of first-order components governed by Eq. (31) where n(E) should be replaced by $n_{ap}(E)$ and N by $N(E) \equiv \int_E^{E_{\max}} n_{ap}(s)ds$.

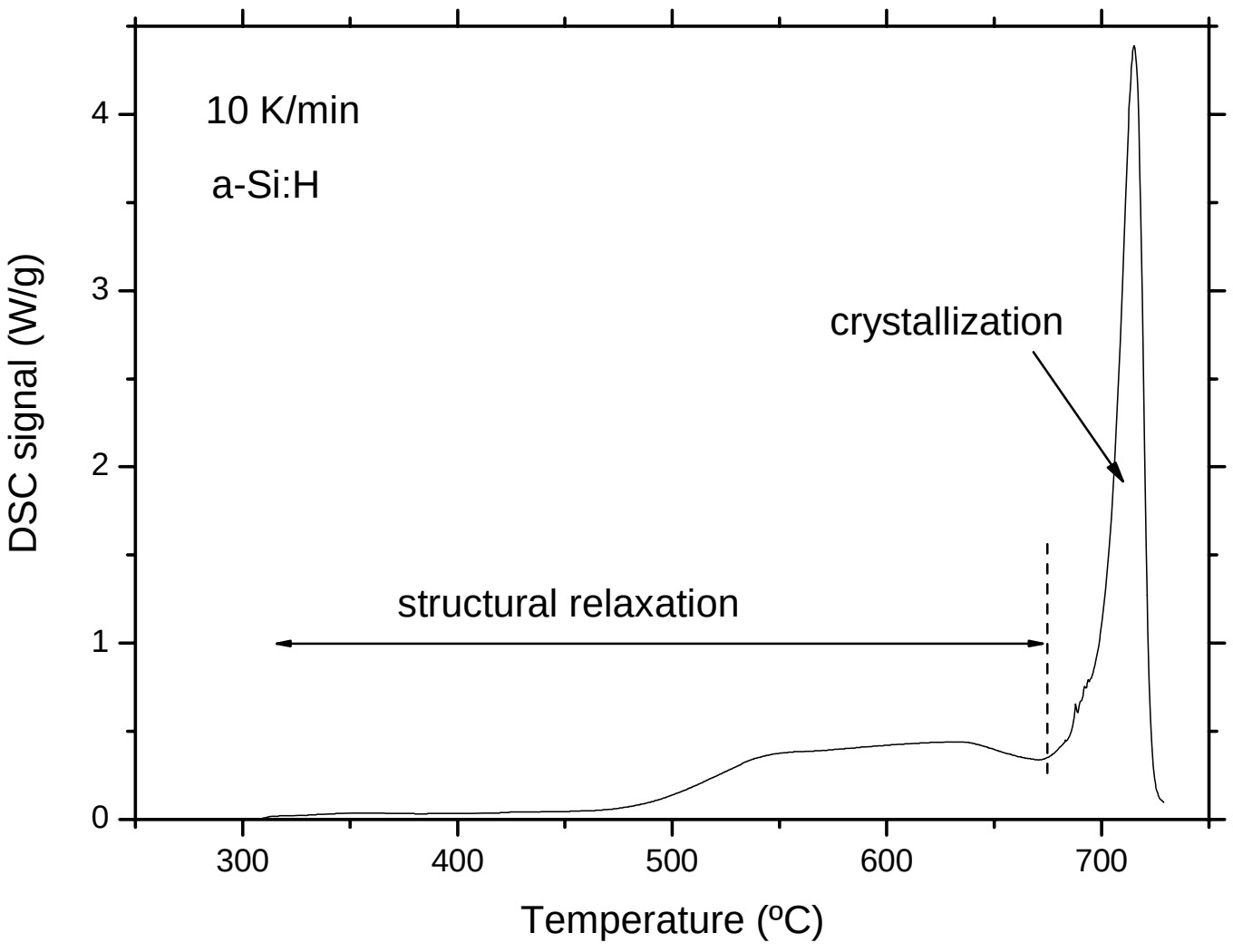

Fig. 1.- Typical DSC thermogram measured in hydrogenated amorphous silicon showing the relaxation and crystallization signals.

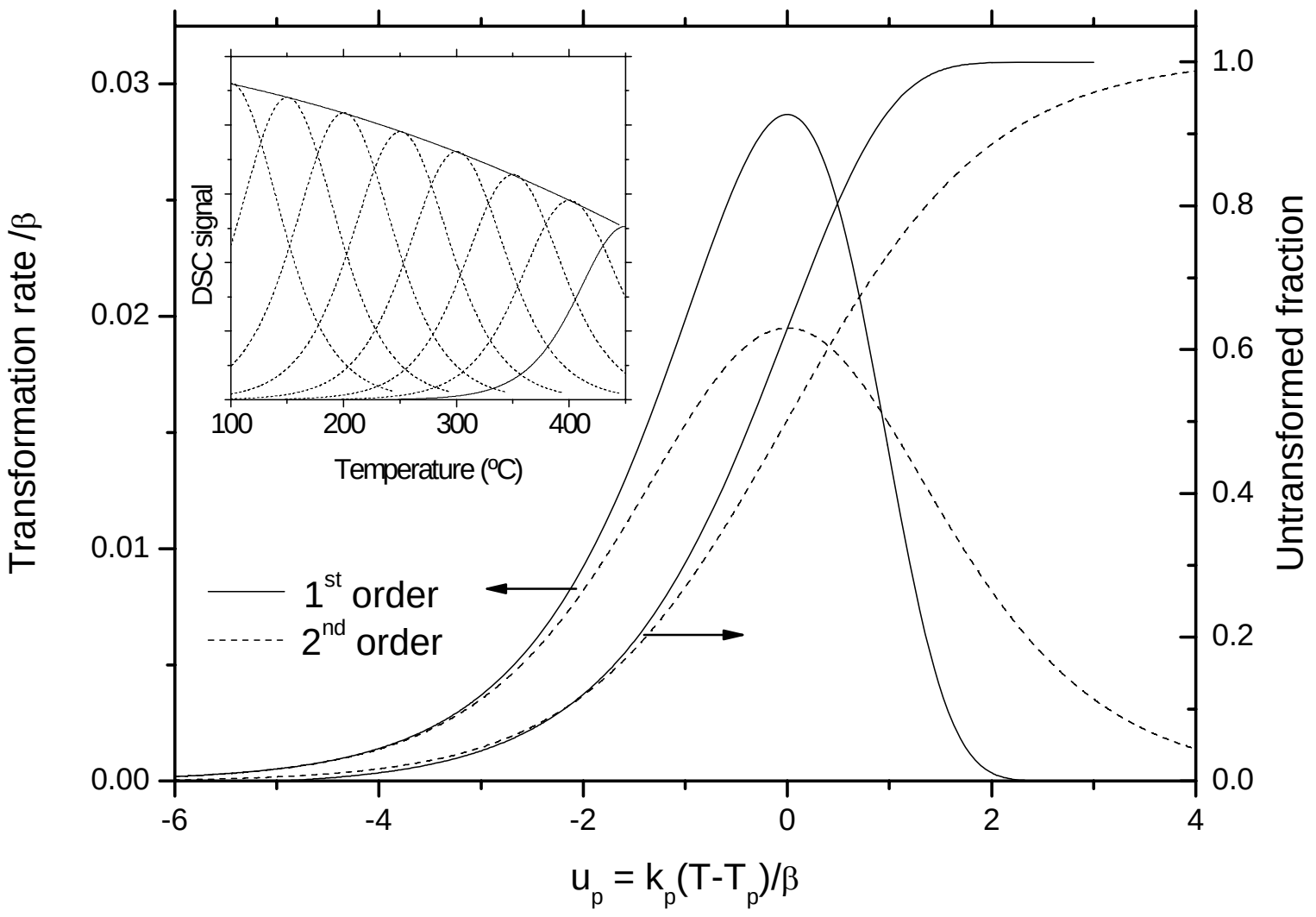

Fig. 2.- Transformation-rate peak-shapes for first- and second-order processes and the corresponding untransformed fractions. Inset: sketch of the contribution of a continuum of components to the DSC signal.

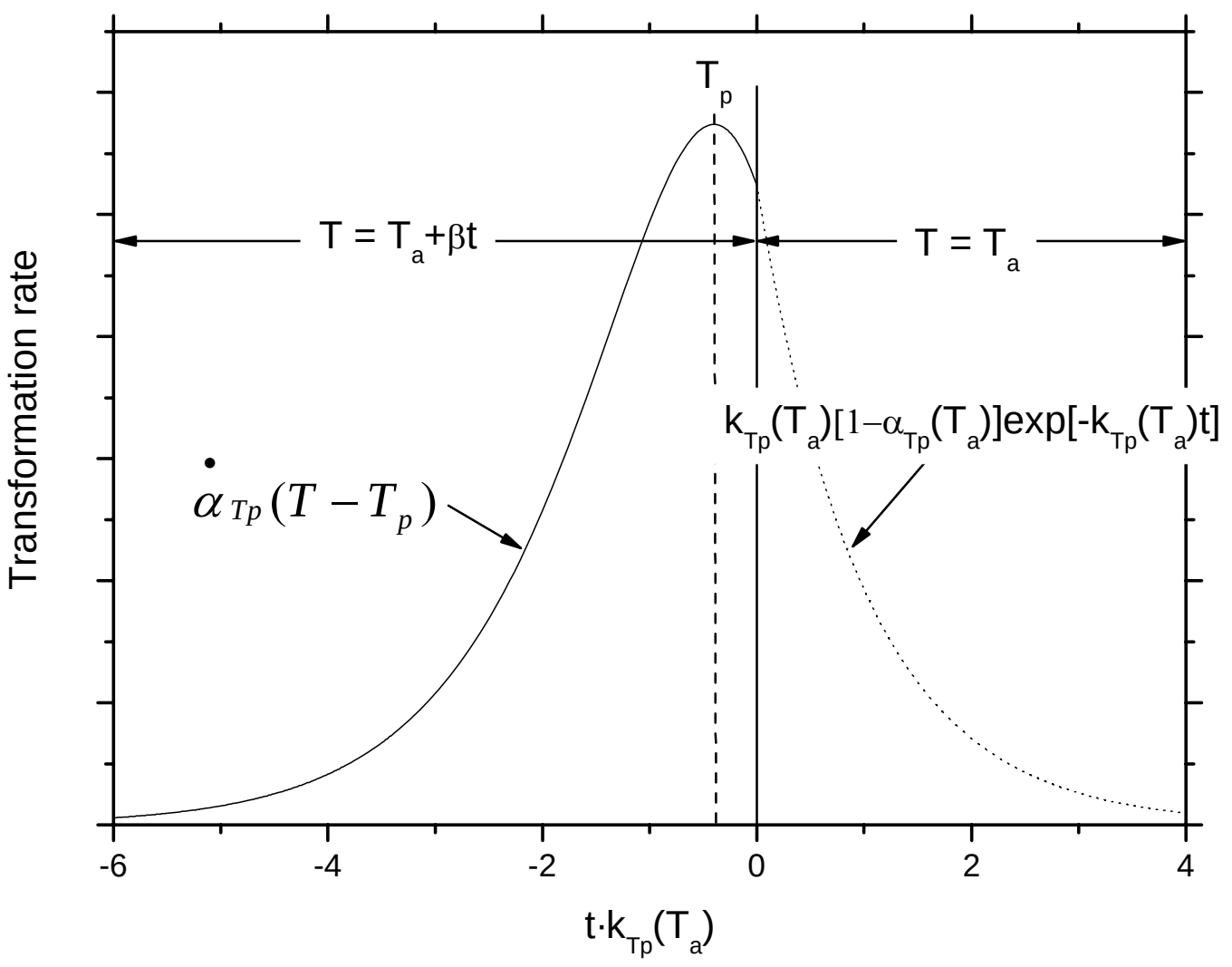

Fig. 3.- Evolution of the transformation rate for a first-order component when an isothermal period is reached.

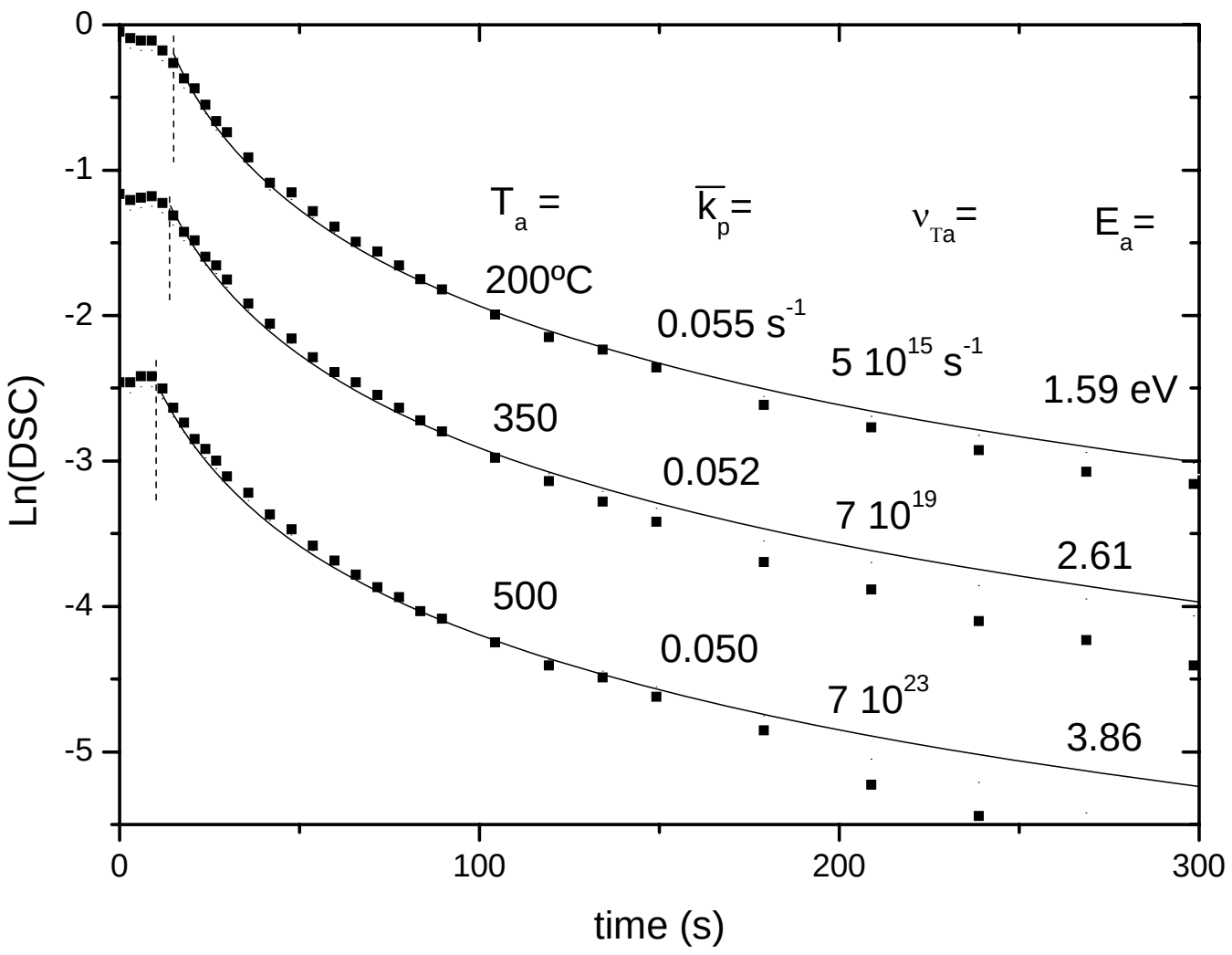

Fig. 4.- Decay of DSC signal when the heating ramp (at 40 K/min) reaches an isothermal period: points (experimental [19]), lines [best fits to Eq. (21)]. Vertical lines indicate the beginning of the isotherm. In the original reference [19], the temperatures of 200 and 500ºC were erroneously exchanged.

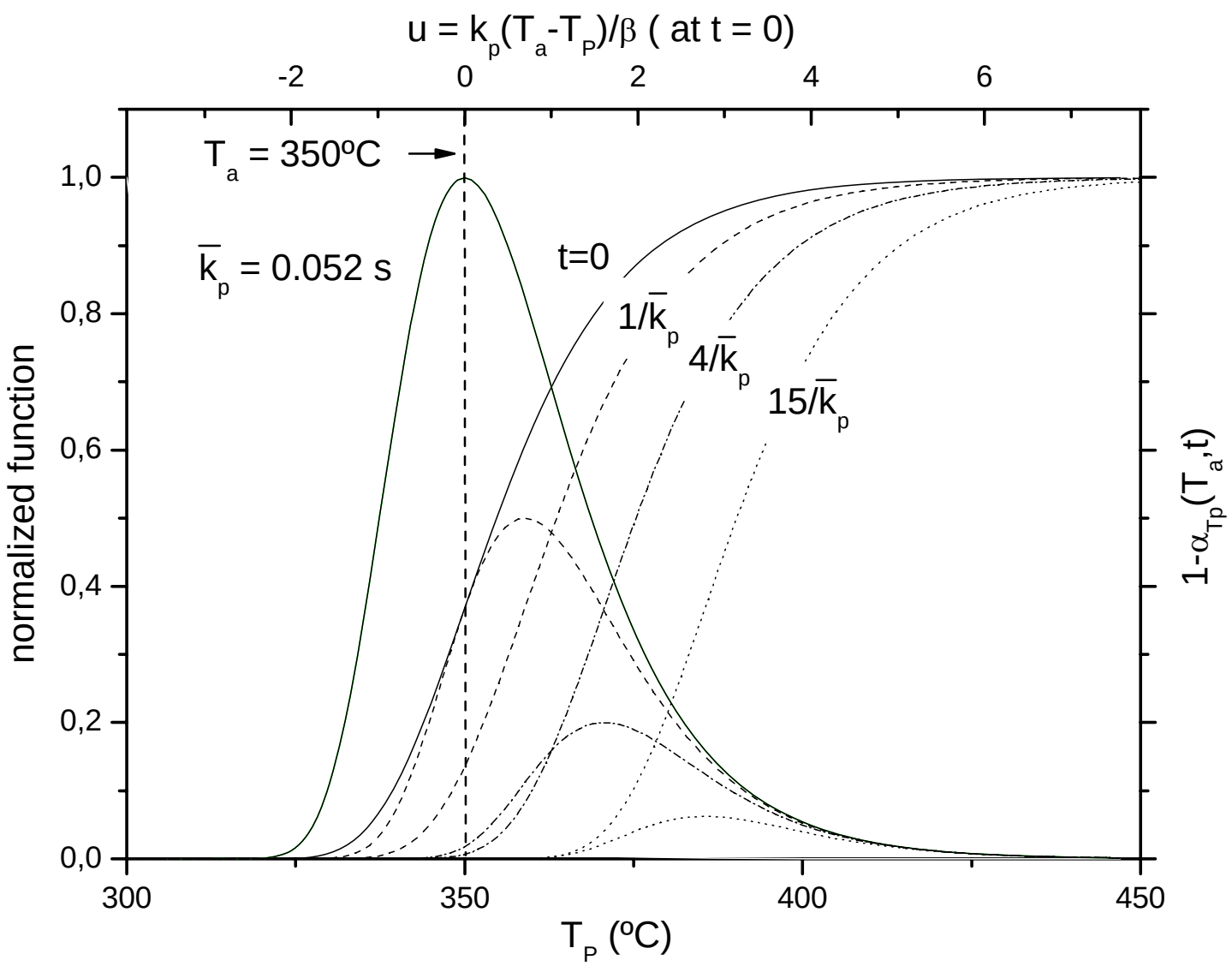

Fig. 5.- Contribution of the different $T_p$ components to the DSC signal measured during isothermal decay at various annealing times (peaks) and the untransformed fraction of these components (sigmoidal curves). ). The $\overline{k_p}$ and $T_a$ values correspond to one experiment on a-Si of Fig 4.

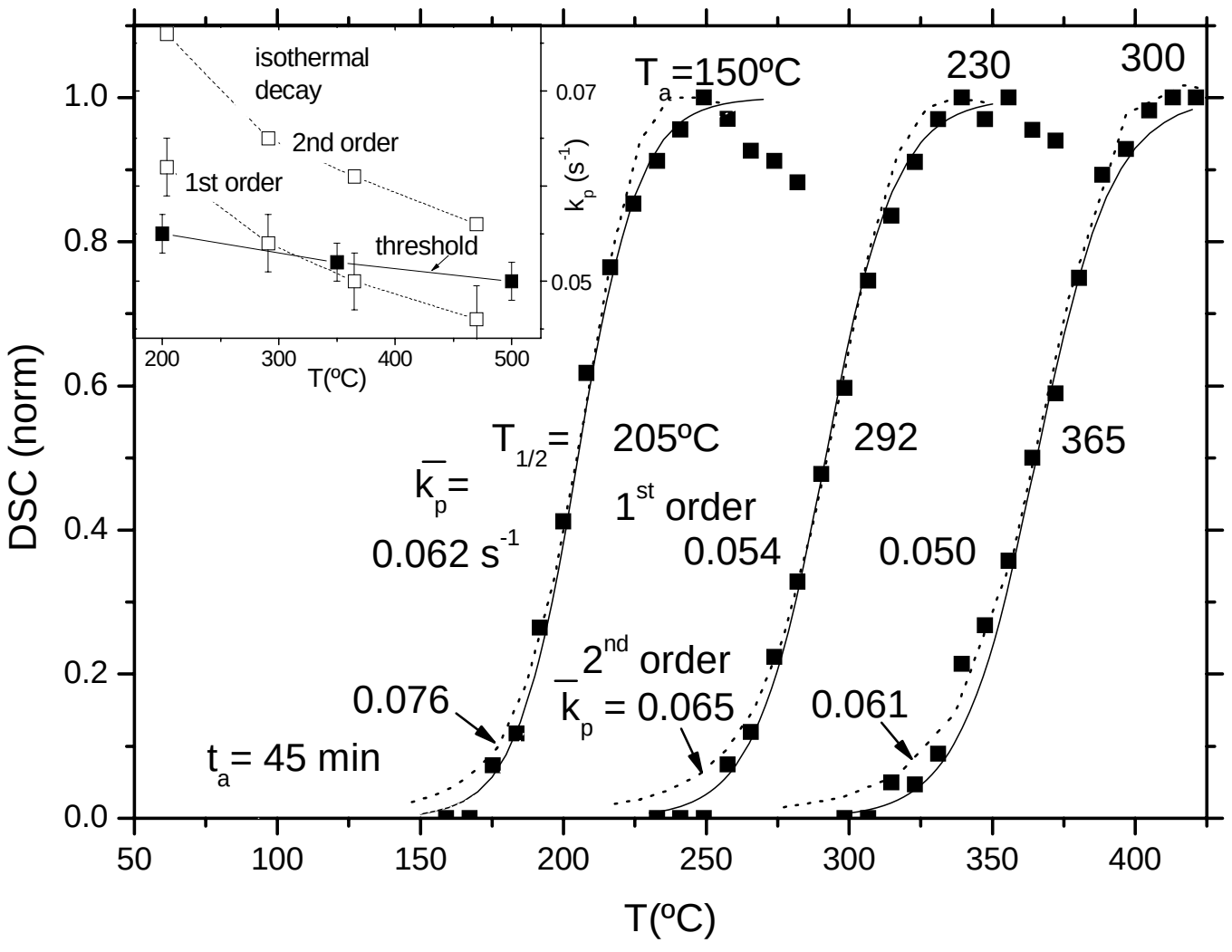

**Fig. 6.-** Relaxation threshold of the DSC signal after a preannealing treatment lasting 45 min at $T_a$: points (experimental results [19]), solid lines [best fits to Eq. (22) for 1st order processes], dotted lines (best fits to 2nd order processes). Inset: comparison of the $\overline{k_p}$ values obtained from the relaxation threshold and from the isothermal decays. Since digitalization of the weak DSC signal after the annealing at 400ºC was not accurate enough, the fourth open point was obtained from the $T_{1/2}$ value [Eq. (27)].

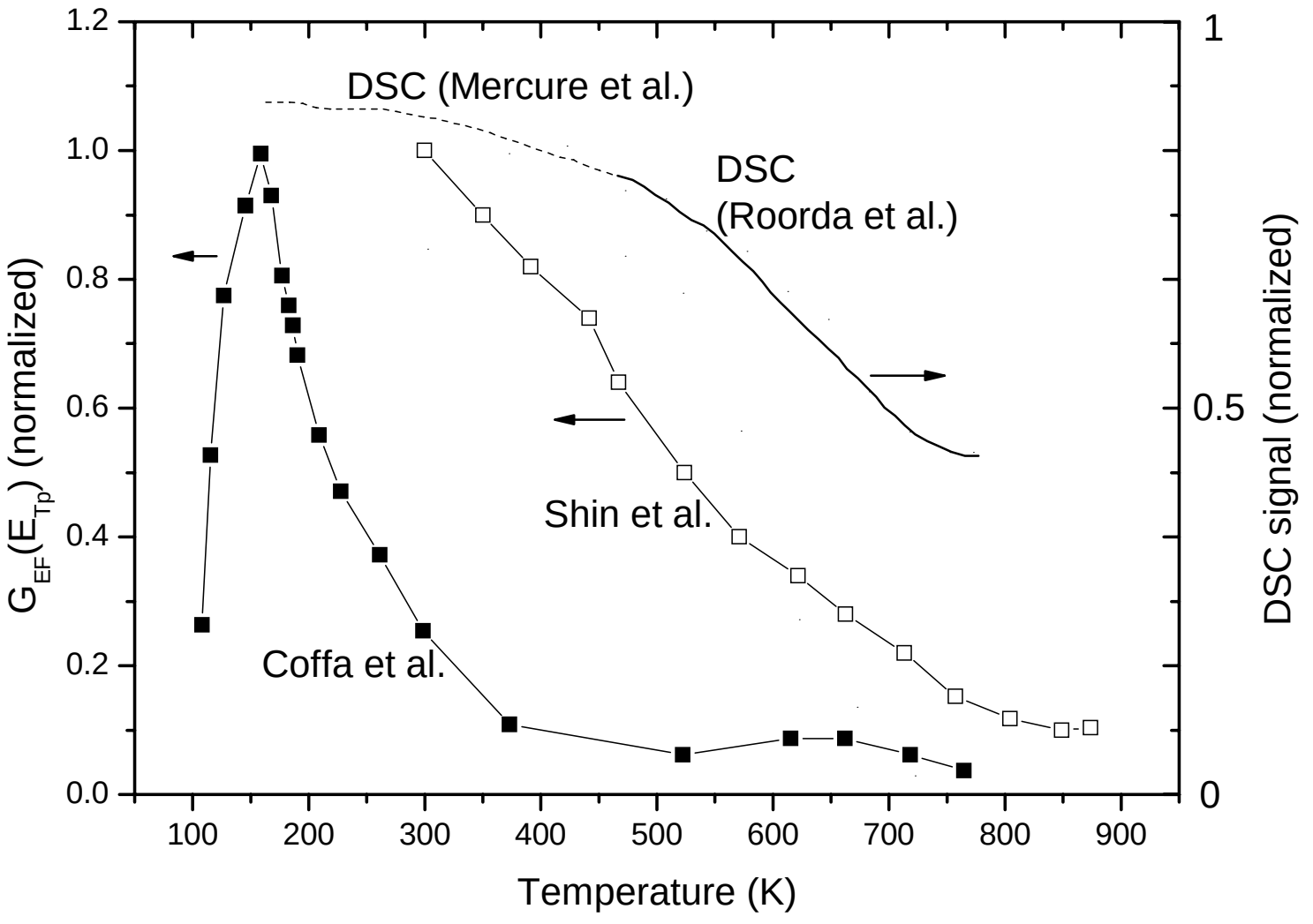

Fig. 7.- Comparison of the density of defects that recombine at a given temperature with the DSC signal measured by several authors.

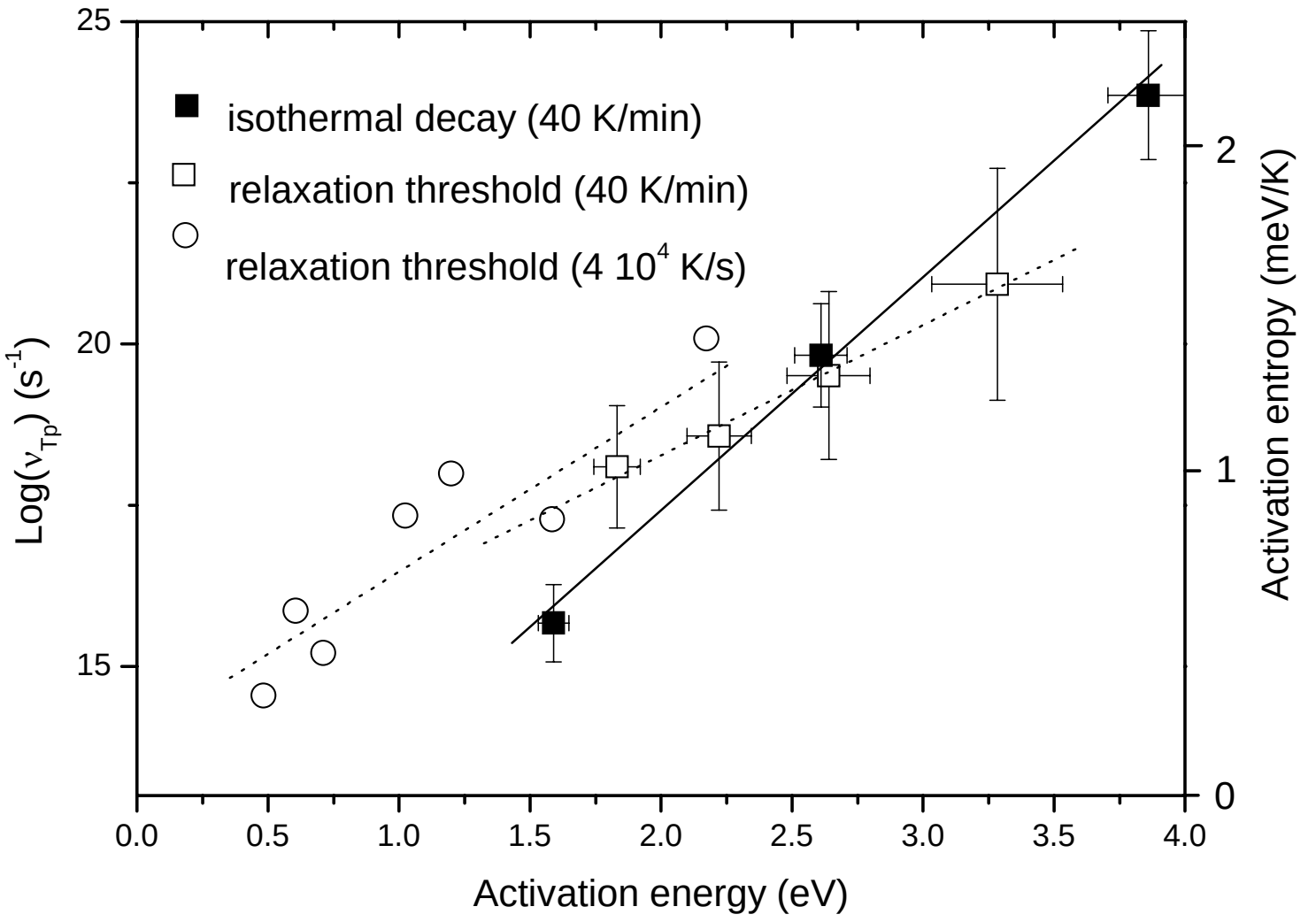

Fig. 8.- Kinetics parameters obtained from the fitting to a superposition of first-order processes of the experimental results of refs. [17] (squares) and [19] (circles).

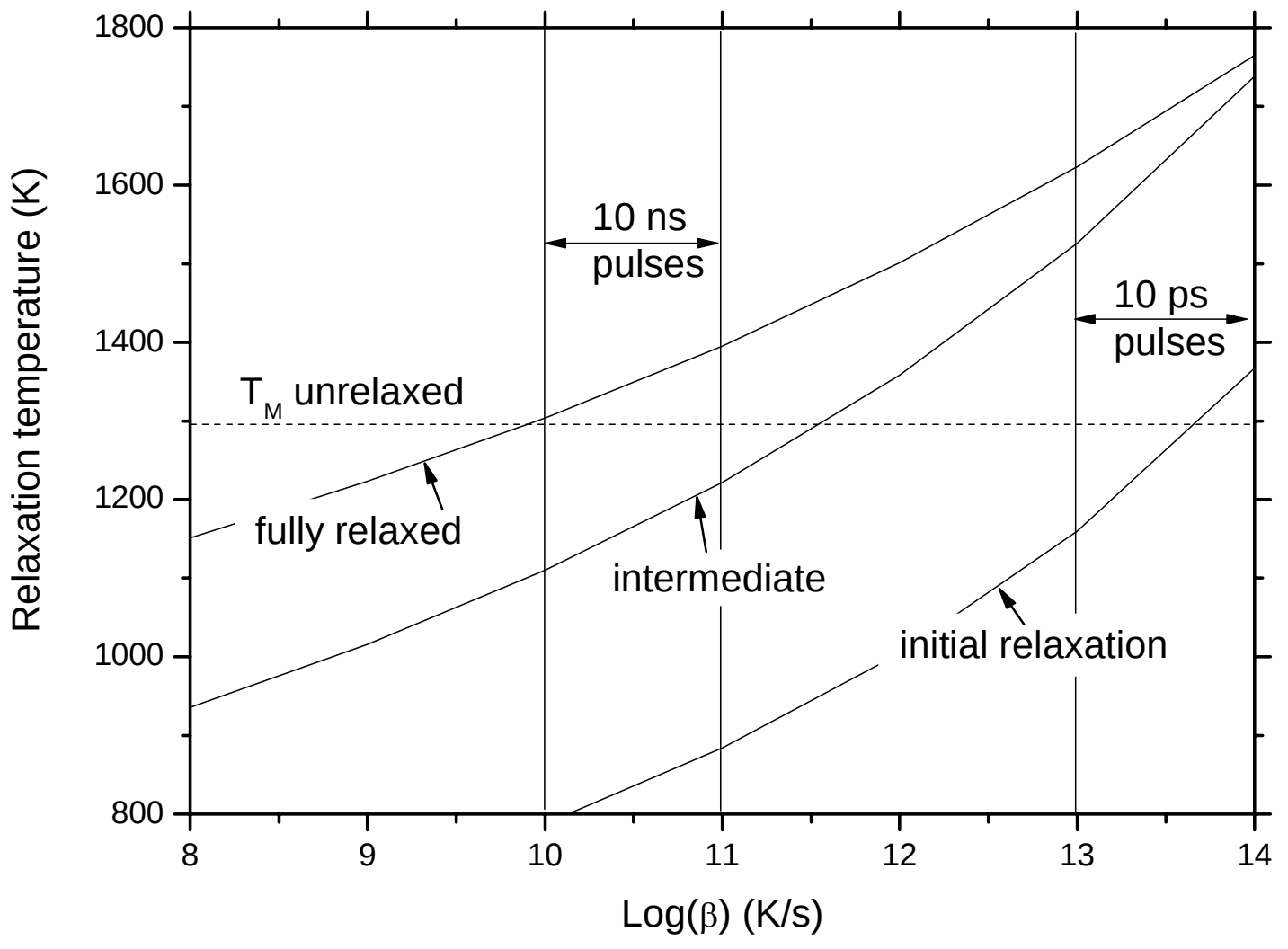

Fig. 9.- Comparison of the relaxation temperatures for several degrees of relaxation corresponding to the components that, at 40 K/min, relax at 100 (“initial”), 300 (“intermediate”) and 500ºC (“fully relaxed”) with the melting temperature $T_M$ of unrelaxed a-Si in the region of heating rates achieved with laser pulses.

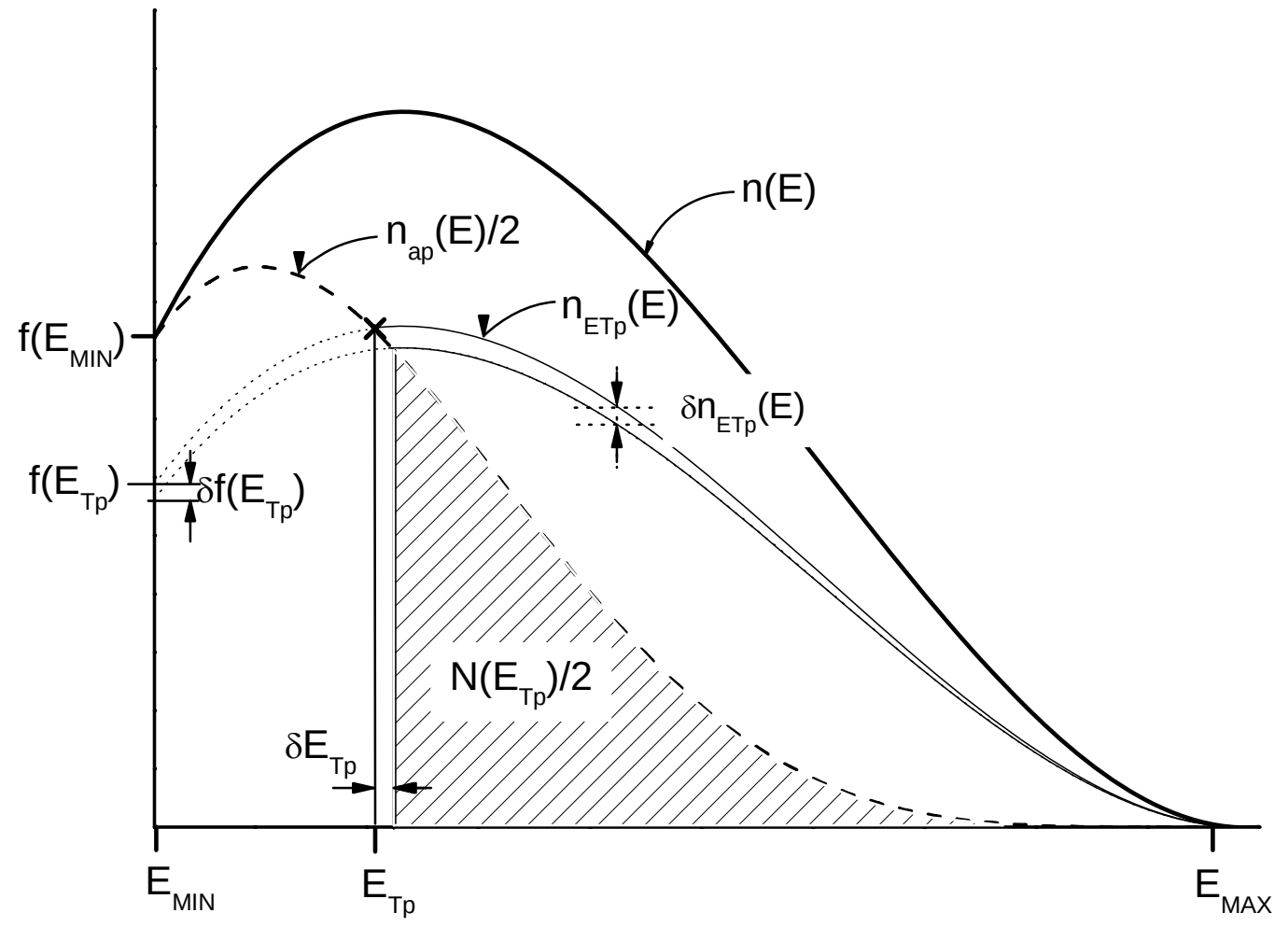

Fig. 10.- The initial distribution of defects, n(E), evolves with time as the mobile defects (defects with low activation energy) disappear. $n_{ETp}(E)$ is the distribution when all defects with $E < E_{Tp}$ have recombined. $n_{ap}(E)$ is the apparent distribution measured.